\documentclass[10pt, a4paper]{article}
\usepackage{graphicx}
\usepackage{amsthm,amssymb,color}
\usepackage{amsmath}
\usepackage{array}
\allowdisplaybreaks
\usepackage{float}
\usepackage{setspace}
\usepackage{multirow,stackengine}
\usepackage{booktabs}
\usepackage{hyperref}

\date{}

\newtheorem{theorem}{Theorem}[section]

\newtheorem{example}[theorem] {Example}

\newcommand{\p} {\mathbb{E}(K)}
\newcommand{\q} {1 - \mathbb{E}(K)}
\newcommand{\tht} {\mathbb{E}(W)}
\newcommand{\oneminustht} {1 - \mathbb{E}(W)}
\newcommand{\alp} {\mathbb{E}(V)}
\newcommand{\bet} {1 - \mathbb{E}(V)}
\newcommand{\yet}{W}
\newcommand{\si}{K}
\newcommand{\gs}{{\bf{m}}}
\newcommand{\wx} {\Theta_{}}

\newcommand{\wa}{\varpi}
\newcommand{\az}{\Gamma_{1}}
\newcommand{\bz}{\Gamma_{2}}
\newcommand{\ftilde}{\widetilde{f}}
\newcommand{\gtilde}{\widetilde{g}}
\newcommand{\wxtilde}{\widetilde{\wx}_{X+\yet Y}}
\newcommand{\wxauxtilde}{\widetilde{\wx}_{\si \left( X +\yet Y \right) + \hat{Y}}}
\newcommand{\mauxtilde}{\widetilde{\gs}^{aux}}
\newcommand{\mtilde}{\widetilde{\bf{m}}}
\newcommand{\s}{\;}

\begin{document}
\title{\begin{center} On the Compound Beta-Binomial Risk Model  with Delayed Claims and Randomized Dividends \end{center}
}
\author{
{\bf  Aparna B. S, Neelesh Upadhye } \\
 \\
}

\maketitle 
\begin{abstract}  
In this paper, we propose the discrete time Compound Beta-Binomial Risk Model with by-claims, delayed by-claims and randomized dividends. We then analyze the Gerber-Shiu function for the cases where the dividend threshold $d=0$ and $d>0$ under the  assumption that the constant discount rate $\nu \in (0,1)$. More specifically, we study the discrete time compound binomial risk model subject to the assumption that the probabilities with which the claims, by-claims occur and the dividends are issued are not fixed(constant),  instead the probabilities are random and  follow a Beta distribution with parameters $a_{i}$ and $b_{i}$, $i = 1, 2, 3$. Recursive expressions for the Gerber-Shiu function corresponding to the proposed model are obtained. The recursive relations are further utilized to obtain significant ruin related quantities of interest. Recursive relations for probability of ruin, the probability of the deficit at ruin, the generating function of the deficit at ruin and the probability of surplus at ruin and for the probability of the claim causing ruin are obtained.
\end{abstract}
{{\it Key words}: Compound Beta Binomial Risk model, Utility function, Gerber-Shiu funciton, Delayed claims, Randomized Dividends.}
\section{Introduction}  
Risk theory is a mathematical construct that describes the vulnerability of a company towards ruin. Ruin related quantities of interest such as probability of ruin, the generating function of the deficit at ruin, the distribution of the surplus etc. may be evaluated using the concept of risk theory. The capital begins with an initial surplus amount $u$, and continues to increase with periodic premiums. The capital decreases in jumps due to claims. Ruin/Dissolution follows whenever the company's reserve amount or surplus becomes negative. Many of the abstract risk models use continuous time risk models but the realistic scenario is the opposite. The advantage of discrete time risk models is that recursive formulae may be arrived at without assuming the distritution of the claim sizes, thereby making it computationally easy. The findings from discrete time risk models serve as simpler forms of their continuous time analogues and may be used as approximations and bounds for the corresponding results in continuous time risk models, see, for example, \cite{dickson1995some}. Related literature on discrete time risk models can be found in ~\cite{xiao2007compound}, ~\cite{liu2015discrete}, ~\cite{eryilmaz2014distributions},\cite{yuen2001ruin},~\cite{tan2006compound}, ~\cite{wat2018compound}, ~\cite{yuen2005ultimate}. For continuous time risk models involving delayed claims, the interested reader is reffered to ~\cite{dassios2013risk}, for an extension of the compound Poisson risk model. In addition, ~\cite{zou2010ruin} involves results having the claim number process as an Erlang(2) process. 
\noindent
The Compound Binomial Risk model, first introduced by Gerber \cite{gerber1988mathematical} in the year 1988, has a utility function given by 
$$\mathbb{S}_{n}= \mathbb{S}_{0} + n - \sum_{i=1}^{N_n} X_i = \mathbb{S}_{0} + n -\sum_{i=1}^{n}K_{i} X_{i},$$
where $\mathbb{S}_{0}= u$ denotes the insurer's initial surplus and $u \in {\mathbb N}$, $n$ represents the discrete time units, the periodic premium rate is assumed to be one unit. Also, $\{X_{i}\}_{i=1}^n$ are independent and identically distributed (iid) random variables (r.v's), where each $X_i$ denotes the size of a main claim at time $i$ having pmf $P(X = k) = f(k)$, $N_n$ represents the number of occurrences of main claims in $n$ time periods, which follows Binomial distribution with parameters $n$ and $\Lambda_{1}$, and is assumed to be independent of $X_i$. Equivalently, if  $K_{i}$ denotes Bernoulli($\Lambda_{1}$) r.v representing the occurrence of a main claim at time $i$, then it is clear that $N_n = K_1+\ldots+K_n$ which implies $\sum_{i=1}^{N_n} X_i = \sum_{i=1}^{n}K_{i} X_{i}$. 
\noindent
In practice, claims can be split into two categories, main claims and by-claims. By-claims are induced by main claims. By-claims are settled in the same time period or at most one time period later. 
\noindent
Let $Y_{i}$ denote the size of the by-claim at time $i$ having pmf $P(Y = k) = g(k)$. Further, $\{Y_{i}\}_{i=1}^n$ be a sequence of iid r.v's representing by-claim sizes at various time periods $i$ having a common pmf  $g(k)$ for $k = 1, 2, 3, \cdots$.  Also, let $W_{i}$ be Bernoulli($\Lambda_{2}$) r.v representing the occurrence of a by-claim at time $i$. We assume that the r.v's $K_{i}$, $W_{i}$, $X_{i}$ and $Y_{i}$ are all independent of each other for all $i$.
 Then the modified utility function is given by 
$${\mathbb S}_{n} = {\mathbb S}_{0} + n - \sum_{i=1}^{n}K_{i}(X_{i} + W_{i}Y_{i}), ~~~{\rm with}~~~ {\mathbb S}_0 = u.$$ 
\noindent     
 This model is further generalized to accommodate randomized dividends and by-claims. 
\noindent
A dividend is due for payment whenever the surplus exceeds or is equal to a threshold $d$, (a pre-determined constant). Let $V_i$ be Bernoulli($\Lambda_{3}$) r.v representing the payment of unit dividend at time $i$. Thus, the modified utility function   is given by 
\begin{align} \label{model0}
\mathbb{S}_{n} ={\mathbb S}_0 + n - \left\{ \sum_{i=1}^{n} V_{i} \mathbb{I}({\lbrace \mathbb{S}_{i-1}\geq d \rbrace}) + \sum_{i=1}^{n-1} K_i(X_i+Y_i) + K_n (X_n + W_n Y_n) \right\}
\end{align}
where $\sum_{i=1}^{n}V_{i} \mathbb{I}({\lbrace \mathbb{S}_{i-1}(u)\geq d \rbrace})$ denotes the total number of dividends paid in $n$ time periods and 
\noindent
The utility function ${\mathbb S}_n$ defined in (\ref{model0}), with $n =0, 1,2,\ldots $ is also known as the discrete time surplus process (DTSP). The time of ruin $\tau$ is defined as  $\tau =\inf \lbrace n > 0 ~|~ \mathbb{S}_{n}<0\rbrace$, the first time that the surplus $\mathbb{S}_{n}$ is negative. The ultimate ruin probability is defined as $\phi(u) =P\lbrace \tau < \infty | \mathbb{S}_{0} = u \rbrace$.
\noindent
At the time of ruin $\{\tau < \infty\}$, $|{\mathbb S}_{\tau}|$ denotes the size of lack/deficit in surplus, ${\mathbb S}_{\tau-} = {\mathbb S}_{\tau - 1}$ is the surplus immediately before ruin and $\mathbb{S}_{\tau -}+|\mathbb{S}_{\tau}|  $ is the  claim causing ruin. Let $\nu \in \left( 0, 1\right]$ be a constant discount rate and $\varpi\left(\mathbb{S}_{\tau-},|\mathbb{S}_{\tau}|\right)$ be a non-negative bounded function, the penalty function which accounts for the gain/loss incurred due to ruin. $\mathbb{I}(\{\tau < \infty\}) $ emphasizes that penalty function is operational only at ruin. The expected penalty with ${\mathbb S}_0 = u$ at ruin is given by Gerber-Shiu discounted penalty function (see, Gerber et al. \cite{gerber1998time}), whenever $d > 0$ has the form $\colon$
\begin{equation}
{\bf m}_{d}\left(u\right)=\mathbb{E}\left[\nu^{\tau}\varpi\left(\mathbb{S}_{\tau-},|\mathbb{S}_{\tau}|\right)\mathbb{I}(\{\tau < \infty\})| \mathbb{S}_{0} = u \right].
\label{Gerbershiu0}
\end{equation}
Since various ruin related quantities of interest may be obtained by choosing the penalty function appropriately, the Gerber-Shiu function has become a significant and standard risk measure in financial literature. Whenever the dividend threshold $d=0$, we denote the corresponding Gerber-Shiu function  $\gs(u)$. For related literature on the Gerber-Shiu function, the interested reader is referred to ~\cite{bao2007gerber}, ~\cite{dong2012complete},~\cite{yu2018absolute}, ~\cite{yu2013some}, ~\cite{willmot2003gerber}. \\
\noindent
In the current paper, we consider a generalization of the Compound binomial risk model with by-claims and randomized dividends wherein the probabilities of occurrence of main claims, by-claims and issuance of dividends, follow a beta distribution. In addition to adopting the model as in \cite{yuen2013discrete} and \cite{wat2018compound}, we also include the assumption that a randomized dividend of 1 with probability $\Lambda_{3}$ is paid until the insurance company goes into ruin. We derive explicit recursive expressions for the discounted Gerber-Shiu function. 
The remaining sections of the paper is organized as follows : In Section 2, we describe the motivation for the model chosen in this paper and define the Beta-binomial compound binomial risk model with by-claims and  randomized dividends, analyze the model and derive explicit expressions for the discounted expected penalty function or the Gerber-Shiu function. In Section 3, the outcomes achieved in Section 2 are utilized to analyze some ruin related quantities of interest namely$\colon$ the probability of ruin, probability of the deficit at ruin, the generating function of the deficit at ruin and the probability of the surplus at ruin.
\section{Compound Beta-Binomial Risk Model and Recursive Relations}
To the best of our knowledge, various DTSPs studied in literature assume that the probabilities  of occurrence of claims, by-claims and the probability of issuance of dividend  are fixed (constant), which is not always a constructive assumption. In practice, the claim probabilities and the dividend issuance probability,  can be random and may have some distribution on $[0,1]$. In order to comprehend the behavior of the DTSP with random claim probabilities we propose the new model with the following assumptions along with the assumptions already made for $(\ref{model0})$.
\begin{itemize}
\item[(A1)] Let $K_i$ be Bernoulli($\Lambda_1$) r.v's that represent the occurrence of a main claim at time $i$, where  $\Lambda_1$ has Beta distribution with parameters $(a_{1},b_{1})$. Hence  $P(K =1) = \mathbb{E}(K)$.
\item[(A2)] Let $W_i$ be Bernoulli($\Lambda_2$) r.v's that represent the occurrence of a by-claim at time $i$, where $\Lambda_2$ has Beta distribution with parameters $(a_{2},b_{2})$. Hence $P(W =1) = \mathbb{E}(W)$
\item[(A3)] Let $V_i$ be Bernoulli($\Lambda_3$) r.v's that represent the payment of unit dividend at time $i$, where $\Lambda_3$ has Beta distribution with parameters $(a_{3},b_{3})$. Hence $P(V =1) = \mathbb{E}(V)$
 \item[(A4)] The expected positive security loading condition (see \cite{damarackas2015note}) for the utility function in DTSP(\ref{model0}) being $1-\alp > \p  ~ \mathbb{E}\left(X+Y\right)$.
 \item[(A5)] $K_{i}$, $W_{i}$, $V_{i}$, $X_{i}$ and $Y_{i}$ are independent of each other for all $i \in \mathbb{N}$.
\end{itemize}
\noindent
Henceforth, the DTSP (\ref{model0}) subject to the assumptions (A1) through (A5),  in addition to the assumptions already made for the Compound Binomial  risk model is called the Compound Beta-Binomial risk model with delayed claims and randomized dividends. In this paper, we analyze the expected discounted penalty function (\ref{Gerbershiu0}) with and without the positive dividend threshold/barrier $d$  for the risk model defined in DTSP$(\ref{model0})$.
\noindent
 We denote the probability generating functions of $f$ and $g$ are given by $\widetilde{f}(z) = \displaystyle\sum_{u=0}^{\infty} z^{u} \s P(X=u)$ and  $\widetilde{g}(z) = \displaystyle\sum_{u=0}^{\infty} z^{u} \s P(X=u)$ respectively. If $ P(X+Y = k) = f*g(k)$,    then $\widetilde{f*g}(z)= \widetilde{f}(z) \widetilde{g} (z)$.
Suppose that a main claim occurring with probability $\Lambda_{1}$ induces a by-claim with probability $\Lambda_{2}$ and the settlement of the by-claim is done simultaneously or in the next time period with probability $1- \Lambda_{2}$, then such a by-claim is called the deferred/delayed by-claim. Assuming that the delayed by-claim occurs in the first time period, the auxiliary utility function or the auxiliary discrete time surplus process (ADTSP) in case of delayed by-claims is given by 
\begin{equation} \label{model2}
\mathbb{S}^{aux}_{n}(u)=u+n-\displaystyle\sum_{i=1}^{n}V_{i} \mathbb{I}{\lbrace \mathbb{S}_{i-1}(u)\geq d \rbrace}- \displaystyle\sum_{i=1}^{n-1} K_{i}\left(X_{i}+W_{i} Y_{i}\right)- K_{n}\left(X_{n}+W_{n}Y_{n}\right) - \hat{Y}  \mathbb{I}_{\lbrace n \geq 1 \rbrace}
\end{equation}
Here, $\hat{Y}$ is a r.v representing the delayed by-claim and having pmf $P(\hat{Y}=k)=g(k)$.  Further, $\hat{Y}$ has the same distribution as $Y$ 
 i.e $\hat{Y}\stackrel{d}{\ =} Y$ . $\hat{Y}$ is independent of all other r.v's. The reader is reffered to  \cite{yuen2005ultimate},  $\cite{yuen2013discrete} $ and $\cite{wat2018compound}$ for literature on risk models with delayed claims. We denote the Gerber-Shiu function corresponding to the ADTSP by  $\gs^{aux} (u)$.
\noindent
Given $Z_{\tau}$ and $P(Z_{\tau } = k)$, the conditional expected penalty function when the DTSP $(\ref{model0})$ becomes negative (from $\tau =u+1 $ up to $\infty$ ) is given by $\colon$
$ \Theta_{Z_{\tau}}(u) = \mathbb{E}\left[\s \varpi\left(\mathbb{S}_{\tau -}, |\mathbb{S}_{\tau}|\right)| \s {\tau < \infty},\mathbb{S}_{\tau-} = u, Z_{\tau } = X \right] $ then,
$\mathbb{E}[\Theta_{Z_{\tau}}(u)]= \displaystyle\sum_{k=u+1}^{\infty}\varpi\left(u,k-u \right) P\left( X=k\right) $. 
We now derive explicit expressions for $\gs(u)$ and ${\gs_{d}}\left(u\right)$ in the first time period which constitutes one of the main results in this paper.
In the first time period, the DTSP $\left(\ref{model0}\right)$ and the corresponding components of the Gerber-Shiu function for various scenarios of claims, by-claims and dividends are listed in Table (1). For example, when  $\tau = 1$, $K = 1$, $W = 1$, $V = 1$, 
$ \mathbb{S}_{1} = \mathbb{S}_{0} - X_{1} - Y_{1}, 
= u  - X_{1} - Y_{1} $, $\mathbb{S}_{0} = u$, a component of $(\ref{Gerbershiu0})$ is given by
\begin{align} \label{first component of m(u)}
& \sum_{k= 0}^{ \infty}\mathbb{E}\left[\s \nu \s \varpi\left(\mathbb{S}_{0},|\mathbb{S}_{1}|\right)| \s \mathbb{S}_{0} = u,  \mathbb{S}_{1} = u  - X_{1} - Y_{1}, K = 1, W = 1, V = 1\right] P(K W V = 1)P(X+Y = k) \nonumber \\ \nonumber
&=  \sum_{k= 2}^{ \infty}\mathbb{E}\left[\s \nu \s \varpi\left(u,|k-u|\right)| \s \mathbb{S}_{0} = u,  \mathbb{S}_{1} = u  - k, K = 1, W = 1, V = 1\right]  P(K W V = 1)  P(X+Y = k) \nonumber \\ 
& = \nu \s \biggl [\sum_{k= 2}^{ u } {\bf m}(u-k)(f * g)(k) +  \sum_{k= u+1}^{ \infty } \varpi(u,k-u) (f *g )(k) \biggr] \mathbb{E}(K)\mathbb{E}(W)\mathbb{E}(V)
\end{align}
The first term in $(\ref{first component of m(u)})$ is for the case when the company does not go into ruin while the second term is for the case when the company goes into dissolution/ruin. Collating the results listed in Table (1) and using the law of total expectation, we may obtain an explicit expression for the expected discounted penalty function $(\ref{Gerbershiu0})$.

 \setlength{\tabcolsep}{2pt}
\begin{table}[H]
 \sffamily\centering
\caption{Components of the Gerber-Shiu function corresponding to DTSP}
 \label{tab:table1}
\begin{center}
\begin{tabular}{|c|c|c|c|l|l|}
\hline
 $K$ & $V$ & $W$ & $S_{1}$ & \multicolumn{2}{c|}{m(u)}  \\
\cline{5-6}
  & & & &  Case of no ruin. & Case of no ruin \\
 \hline
 1 & 1 & 1 & $u-X-Y$ & $\displaystyle\sum_{k= 2}^{ u } \gs(u-k) \s (f*g)(k)$ & $\displaystyle\sum_{k= u+1}^{ \infty} {\varpi}(u,k-u) \s (f*g)(k)$ \\
 1 & 1 & 0 & $u + 1 - X -Y$ & $\displaystyle\sum_{k= 2}^{ u+1 } \gs(u+1-k)\s (f*g)(k)$ & $\displaystyle\sum_{k= u+2}^{ \infty} {\varpi}(u+1,k-u-1) \s (f*g)(k)$ \\
 1 & 0 & 1 & $u - X$ & $\displaystyle\sum_{k= 1}^{ u } \gs(u-k) \s f(k)$ & $\displaystyle\sum_{k= u+1}^{ \infty} {\varpi}(u,k-u) \s f(k)$ \\
1 & 0 & 0 & $u + 1 - X$ & $\displaystyle\sum_{k= 1}^{ u+1 } \gs(u+1-k) \s f(k)$ & $\displaystyle\sum_{k= u+2}^{ \infty} {\varpi}(u+1,k-u-1) \s f(k)$ \\
 0 & 1 & - & $ u $ & ${\bf m}(u)$ & $-$ \\
0 & 0 & - & $ u+1 $ & ${\bf m}(u + 1)$ & $-$ \\
\hline
\end{tabular}
\end{center}
\end{table}

Now, in  DTSP(\ref{model0}) whenever $d=0$, we have,
\begin{equation} \label{utilit}
\mathbb{S}_{n}= u + n-\sum_{i=1}^{n}V_{i}\mathbb{I}{\lbrace \mathbb{S}_{i-1}(u)\geq 0 \rbrace}- \sum_{i=1}^{n-1}
K_{i}(X_{i}+Y_{i})- K_{n}\left( X_{n}+W_{n}Y_{n}\right)
\end{equation}
We now proceed to find a recursive relation for $\gs(u)$ as described below $\colon$
{\footnotesize
\begin{align} \label{thm1part1}
\gs (u) &= v \s [ \q ] \s  [\bet] \s \gs(u+1) + v \s [ \q ] \s \alp  \s \gs(u) \nonumber \\  \nonumber
&+ v \s \p \s \tht  \biggl[ [\bet ] \s \sum_{k= 2}^{ u+1 } \s \gs (u+1-k) \s (f*g)(k) 
+\alp \s \sum_{k= 2}^{ u } \gs (u-k) \s (f*g)(k) \biggr] \\ \nonumber
&+ v \s  \p \s  \tht  \s \biggl[ [\bet] \s \sum_{k= u+2}^{ \infty } \varpi (u+1, k-u-1) \s (f*g)(k) + \alp \s \sum_{k= u+1}^{ \infty } \varpi (u, k-u) \s (f*g)(k) \biggr] \\ \nonumber
&+ v \s \p \s [ \oneminustht ] \s  \biggl[ [\bet] \sum_{k= 1}^{ \ u+1 } \gs^{aux} (u+1-k) \s f(k)+ \alp \s \sum_{k= 1}^{ u } \gs^{aux} (u-k) \s f(k)\biggr] \\ 
&+ v  \s \p \s [\oneminustht ] \s  \bigg[ [\bet] \sum_{k= u+2}^{ \infty } \varpi (u+1, k-u-1) \s f(k)+ \alp \s \sum_{k= u+1}^{ \infty } \varpi (u, k-u) \s f(k) \bigg]
\end{align}
}
Simplifying further we have,

{\footnotesize
\begin{align} \label{thm1part2}                        
&\biggl[1-v \s [ \q ] \s  \alp \s \biggr] \s \gs(u) - v \s [ \q ] \s [ \bet ] \s \gs(u+1) \nonumber \\ \nonumber
&= v \s  \p \s \tht  \s \Biggl[ [\bet] \s (\gs * f * g )(u+1) 
+  \alp \s ( \gs * f * g )(u) \Biggr] \\ \nonumber
&+ v \s \p \s [ \oneminustht ] \s\Biggl[ [ \bet ] \s (\gs^{aux} * f )(u+1)+  \alp \s ( \gs^{aux} * f )(u) \\ 
&+ v \s \p  \s [ \bet ] \s \mathbb{E} [ {\wx}_{X+\yet Y} (u+1)]  + v \s \p \s \alp \s \mathbb{E} [{\wx}_{X+\yet Y} (u)] 
\end{align}
} 
where $(f*g)(u) =\displaystyle\sum_{k= 0}^{ u }f(u-k)g(k)$.
We now inspect the auxiliary utility function $\mathbb{S}_{n}^{aux}(u)$ i.e ADTSP$(\ref{model2})$ when $d=0$ in the first time period. 
\begin{equation} \label{GS function}
\mathbb{S}^{aux}_{n}=u+n-\displaystyle\sum_{i=1}^{n}V_{i} \mathbb{I}{\lbrace \mathbb{S}_{i-1}(u)\geq 0 \rbrace}- \displaystyle\sum_{i=1}^{n-1} K_{i}\left(X_{i}+W_{i} Y_{i}\right)- K_{n}\left(X_{n}+W_{n}Y_{n}\right) - \hat{Y}  \mathbb{I}_{\lbrace n \geq 1 \rbrace}
\end{equation}
The utility function in ADTSP$(\ref{GS function})$ and the corresponding components of the Gerber-Shiu function in the first time period, using the law of total expectation, are listed in Table (2) $\colon$
 \setlength{\tabcolsep}{2pt}
\begin{table}[h!]
\small\sffamily\centering
\caption{Components of the Gerber-Shiu function corresponding to ADTSP}
 \label{tab:table2}
\begin{center}
\begin{tabular}{|c|c|c|c|l|l|}
\hline
 $K$ & $V$ & $W$ & $\mathbb{S}^{aux}_{1}$ & \multicolumn{2}{c|}{ $\gs^{aux}(u)$ }  \\
\cline{5-6}
  & & & &  Case of no ruin. & Case of no ruin \\
 \hline
 1 & 1 & 1 & $u-X-Y - \hat{Y}$ & $\displaystyle\sum_{m= 3}^{ u } \gs(u-m) \s (f*g*g)(m)$ & $\displaystyle\sum_{m= u+1}^{ \infty} {\varpi}(u,m-u)(f*g*g)(m)$ \\
 1 & 1 & 0 & $u + 1 - X -Y - \hat{Y}$ & $\displaystyle\sum_{m= 3}^{ u+1 } \gs(u+1-m) \s (f*g*g)(m)$ & $\displaystyle\sum_{m= u+2}^{ \infty} {\varpi}(u+1,m-u-1) \s (f*g*g)(m)$ \\
 1 & 0 & 1 & $u - X - \hat{Y}$ & $\displaystyle\sum_{m= 2}^{ u } \gs(u-m) \s (f*g)(m)$ & $\displaystyle\sum_{m= u+1}^{ \infty} {\varpi}(u,m-u)(f*g)(m)$ \\
1 & 0 & 0 & $u + 1 - X - \hat{Y}$ & $\displaystyle\sum_{m= 2}^{ u+1 } \gs^{aux}(u+1-m)\s (f*g)(m)$ & $\displaystyle\sum_{m= u+2}^{ \infty} {\varpi}(u+1,k-u-1) \s (f*g)(m)$ \\
 0 & 1 & - & $ u - \hat{Y} $ & $\displaystyle\sum_{m = 1}^{ u } \gs^{aux}(u - m) \s g(m)$ & $\displaystyle\sum_{m = u+1}^{ \infty} {\wa}(u,m-u) \s g(m)$ \\
0 & 0 & - & $ u+1- \hat{Y} $ & $\displaystyle\sum_{m = 1}^{ u+1 } \gs(u + 1 -m) \s g(m)$ & $\displaystyle\sum_{m = u+2}^{ \infty} {\wa}(u+1,m-u-1) \s g(m)$ \\
\hline
\end{tabular}
\end{center}
\end{table}

Collating the various cases in Table $(\ref{tab:table2})$ and using arguments similar to those which led to $(\ref{thm1part1})$ and $(\ref{thm1part2})$, we can obtain a recursive expression for $\gs^{aux}(u)$ as follows:
{\footnotesize
\begin{align*}
&\gs^{aux}(u) = v \s [\q ] \s [ \bet ] \biggl[ \sum_{m = 1}^{ u+1 } {\bf m}(u + 1 -m) \s g(m)+ \sum_{m = u+2}^{ \infty} {\wa}(u+1,m-u-1)\s g(m)\biggr] \nonumber \\
&+ v \s [\q] \s \alp \biggl[ \sum_{m = 1}^{ u } {\bf m}(u - m) \s g(m)+ \sum_{m = u+1}^{ \infty} {\wa}(u,m-u) \s g(m) \biggr] \nonumber \\
&+ v \s \p \s \tht \biggl[[\bet] \s \sum_{m= 3}^{ u+1 } {\bf m}(u+1-m) \s (f * g * g)(m) 
+ \alp \s \sum_{m= 3}^{ u } {\bf m}(u-m) \s (f* g * g)(m)\biggr]\nonumber \\ \nonumber
&+ v \s \p \s  \tht  \biggl[ [\bet ] \s \sum_{m = u+2}^{ \infty} {\wa}(u+1,m-u-1)(f * g * g)(m) 
+ \alp \s \sum_{m= u+1}^{ \infty} {\wa}(u,m-u) \s (f * g * g)(m)\biggr] \\ \nonumber \\ \nonumber
&+ v \s \p \s [ \oneminustht ] \biggl[ [\bet] \s \sum_{m= 2}^{ u +1 } \gs^{aux}(u+1-m) \s (f * g)(m) + \alp \s \sum_{m= 2}^{ u } \gs^{aux}(u - m) \s (f * g)(m)\biggr] \\ \nonumber \\ 
&+ v  \s \p\s [ \oneminustht ] \biggl[ [\bet] \sum_{m = u+2}^{ \infty} {\wa}(u+1,m-u-1) \s (f * g )(m) + \alp  \s \sum_{m = u+1}^{ \infty} {\wa}(u,m-u) \s (f * g)(m)\biggr]
\end{align*}
}
Simplifying further, we obtain :
{ \footnotesize
\begin{align} \label{thm1part3}
& \gs^{aux}(u) = v \s [ \q ] \biggl[ [\bet] \s (\gs * g)(u+1) + \alp \s (\gs * g)(u)\biggr] \nonumber \\
&+ v \s \p \s \tht  \biggl[ [\bet] \s (\gs * f * g * g)(u+1) + \alp \s (\gs * f * g * g)(u)\biggr] \nonumber \\
&+ v \s \p \s  [\oneminustht]  \biggl[ [\bet] \s (\gs^{aux} * f * g)(u+1) + \alp \s (\gs^{aux} * f * g)(u)\biggr] \nonumber \\
&+ v \s \biggl[ [\bet] \s \mathbb{E}[{\wx}_{\left( X + \yet Y\right)+ \hat{Y}} (u+1)] + \alp \s \mathbb{E}[ {\wx}_{\si{\left( X + \yet Y\right)+ \hat{Y}} }(u)]\biggr]
\end{align} 
}
Using the method of generating function and multiplying both sides of  $\eqref{thm1part2}$ and $\eqref{thm1part3}$ by $z^{u+1}$ and summing over $u$ from $0$ to $\infty$, and using the definition of generating function of $f$ and $g$, we obtain the following expressions $\colon$
From $(\ref{thm1part2})$ we get, 
{ \footnotesize
\begin{align} \label{multiply gcapz here}
&\Biggl[ z - v \s \bigg\{ [\bet] + z \s \alp \bigg\} \biggl[ [\q ]+ \p \tht \ftilde(z) \gtilde(z) \biggr] \Biggr] \s \mtilde(z) \nonumber  \\ \nonumber
&= v \s \p \s [ \oneminustht ] \bigg\{ [\bet] + z \s \alp  \bigg\} \ftilde(z) \s \mauxtilde (z)\nonumber \\ \nonumber
&+ v \s  \p   \bigg\{ [\bet] + z \s \alp \bigg\} \mathbb{E}[ \wxtilde (z) ] - v \s [ \q] \s [ \bet ] \s  \gs (0) \\
&- v \s \p \s [ \bet ] \s \mathbb{E}[ {\wx}_{\left( X + \yet Y\right)} (0)] 
\end{align}
} 
Here,$\sum_{u = 0}^{ \infty } z^{u} {\bf m}(u) = \mtilde(z)$  and
$\sum_{u = 0}^{ \infty } z^{u}{\wx}_{\left( X + \yet Y\right)}= \mathbb{E}[\wxtilde (z)] $ \\
From $(\ref{thm1part3})$ we get,
{ \footnotesize
\begin{align} \label{add eqn with gcapz}
&v \s  \bigg\{ [\bet] + z \s \alp \bigg\}  \biggl[ [ \q ]+ \p \s \tht \ftilde(z) \s  \gtilde(z) \biggr] \gtilde(z) \s \mtilde(z) \nonumber  \\ \nonumber
&= \biggl[ z - v \s \p \s [\oneminustht] \bigg\{ [\bet] + z \s \alp \bigg\} \ftilde(z)\gtilde(z)\biggr] \mauxtilde (z)\nonumber \\ \nonumber
&- v \bigg\{ [\bet] + \alp z \bigg\} \mathbb{E}[\wxauxtilde (z)] + v \s [\bet ] \s \mathbb{E}[ {\wx}_{K( X + \yet Y)+ \hat{Y}}] (0) \\ 
\end{align}  
} 
Multiplying $(\ref{multiply gcapz here})$ by $\gtilde(z)$ and adding it to $(\ref{add eqn with gcapz} )$ gives an expression for $\mauxtilde(z)$,which is then substituted back in expression $(\ref{multiply gcapz here})$ to get, 
{ \footnotesize
\begin{align} \label{eqn for mauxtilde}
&\mauxtilde(z) =  \gtilde(z) \s \mtilde(z) \nonumber \\ \nonumber
&-\dfrac{v}{z} \bigg\{[\bet] + z \s \alp \bigg\} \s \biggl[ \p \s \mathbb{E}[\wxtilde(z)] \s  \gtilde(z) - \mathbb{E}[ \wxauxtilde(z)]\biggr] \nonumber \\
&+ \dfrac{v}{z} \s [ \bet ] \biggl[ [\q] \s \gs(0)  \s \gtilde(z) + \p \s \mathbb{E} [ {\wx}_{\left( X + \yet Y\right)}(0) ] \s \gtilde(z) - \mathbb{E}[ \wxauxtilde(0)]\biggr] 
\end{align}
}   
Substituting  for $\mauxtilde(z) $ from expression $(\ref{eqn for mauxtilde})$ in expression $(\ref{multiply gcapz here})$  and defining $\Gamma_{1}(z) = z - \nu \s \p \s [\oneminustht] \bigg\{ [\bet] + z \s \alp \bigg\}  \ftilde(z) \s \gtilde(z) $ and \\
 $\Gamma_{2}(z) = z - \nu \s \p \s [\oneminustht] \bigg\{ [\bet] + z \s \alp \bigg\} \Bigg\{ [\q] + \p \s \ftilde(z) \s \gtilde(z) \Bigg\}$ , we obtain: 
 
{ \footnotesize
\begin{align} \label{subeqn for mauxtilde in multip by gcapz}
& \bz(z) \s \mtilde(z) = v \s \p \dfrac{1}{z} \bigg\{ [\bet]  + z \s \alp \bigg\} \az(z) \s \mathbb{E}[ \wxtilde(z)] \nonumber \\ \nonumber
&+ \dfrac{\nu^{2}}{z} \s \p \s [\oneminustht] \s \bigg\{ [\bet]  + z \s \alp\bigg\}^{2} \s \ftilde(z) \s \mathbb{E}[{\wx}_{\si \left( X + \yet Y\right) + \hat{Y}}] \nonumber \\
& - \dfrac{\nu}{z} \s  [\bet] \s [\q] \s \az(z) \s \gs (0) \nonumber \\ \nonumber
& - \dfrac{\nu}{z} \s \p \s [\bet] \s \az(z) \s  \mathbb{E}[{\wx}_{ X + \yet Y}(0) ]  \\
& - \dfrac{\nu^{2}}{z}\s \p \s [\oneminustht] \s [\bet] \s \bigg\{ [\bet]  + z \s \alp\bigg\} \s \ftilde(z) \s \mathbb{E}[{\wx}_{\si \left( X + \yet Y\right)+ \hat{Y}} (0)] 
\end{align} 
}
Here $\gs(0)$, the initial value is unknown. It is observed that, since $\ftilde(0) =0$  $\gtilde(0) = 0$, $\ftilde(1) = 1$ and $\gtilde(1) = 1$,  $\ftilde^{'}(1) = \mathbb{E}(X) $, $\gtilde^{'}(1) = \mathbb{E}(Y) $, 
$\bz(0) = - v [\bet ] \s [\q] < 0 $ and $\bz(1) = 1 - v \bigg\{ \bet + \alp\bigg\}\biggl[ [\q] + \p \ftilde(1) \gtilde (1)\biggr] \geq 1- v > 0$. Thus, there exists at least one real root of $\bz(z)$ in the interval $(0,1)$. Let $z_{0}$ be the root of $\bz(z)$ in $(0,1)$. Further, since $\ftilde^{'}(z) \leq \ftilde^{'}(1)$, $\gtilde^{'}(z) \leq \gtilde^{'}(1)$, $\ftilde(z) \gtilde^{'}(z) - \ftilde^{'}(1) \gtilde(1) \leq 0 $ and $\ftilde^{'}(z) \gtilde(z) - \ftilde(1) \gtilde^{'}(1) \leq 0$. So we have, 
{ \footnotesize
\begin{align*}
\dfrac{d}{dz}\bz(z) &=  \Biggl[1 - v \s \alp  \bigg\{ [\q] + \p \ftilde(z) \gtilde (z)  \bigg\} \\ \nonumber
& - v \s  \bigg\{ [\bet] + z \s \alp   \bigg\} \bigg\{ \p \s  \ftilde^{'}(z) \gtilde(z) + \p \s  \ftilde(z) \gtilde^{'}(z) \ \bigg\} \Biggr] \\
&\geq 1 - v \s \alp \s \bigg\{ [\q] + \p \ftilde(1) \gtilde (1)  \bigg\} \\ 
&- v \s  \bigg\{ [\bet] + \s \alp   \bigg\} \bigg\{ \p \s  \ftilde^{'}(1) \gtilde(1) + \p \s  \ftilde(1) \gtilde^{'}(1) \ \bigg\}  \\
&= 1 - v \s \alp - v \s \p \s  [\mathbb{E}(X) + \mathbb{E}(Y)] > 0 \\ 
& \geq 1 - \nu \s \p \s  [\mathbb{E}(X) + \mathbb{E}(Y)]  \s \s \s \s \s \text{(due to the positive security loading condition already assumed.)} \\ 
& \geq 1 - \nu \s \alp - \nu \s [\bet] = 1 - \nu > 0 
\end{align*}
}
Thus, $\bz(z)$ is a strictly increasing function on $(0,1)$. This ensures that $\bz(z)=0$ has a unique root $z_{0}$ in $(0,1)$.
Replacing $z = z_{0}$ in $(\ref{subeqn for mauxtilde in multip by gcapz})$ gives the initial value $\gs(0)$ as $\colon$ 
{ \footnotesize
\begin{align}  \label{initial value mzero}
\gs(0)&= \dfrac{\p}{[\q]} \s\Bigg\{ \biggl[1 + \dfrac{z_{0} \s \alp}{[\bet]}  \biggr] \s \mathbb{E} [\tilde{\wx}_{X+\yet Y}(z_{0})] -  \mathbb{E}[ {\wx}_{X+\yet Y}(0)]  \Bigg\}\nonumber \\ \nonumber
& +\Bigg\{\dfrac{\p}{[\q]} \s \times \dfrac{\nu \s [\oneminustht] \s [\bet] }{\az(z_{0})} \ftilde(z_{0})  \biggl[ 1 + \dfrac{z_{0} \s \alp }{[\bet]} \biggr] \Bigg\} \\
& \times \Bigg\{  \biggl[ 1 + \dfrac{z_{0} \s \alp }{[\bet]} \biggr]  \mathbb{E}[{\wx}_{\si \left( X + \yet Y\right)+ \hat{Y}} (z_{0})] -  \mathbb{E}[{\wx}_{\si \left( X + \yet Y\right)+ \hat{Y}} (0)] \Bigg\}
\end{align} 
}

Comparing the coefficients of $z^{u+1}$ on both sides of $(\ref{subeqn for mauxtilde in multip by gcapz})$  the recursive relation for the expected discounted penalty function $\gs(u)$ for the case when $d = 0$ can be obtained as$\colon$ 
{ \footnotesize
\begin{align} \label{main result}
& \gs(u+1) = \biggl[ \nu  \s [\q] \s [\bet] \biggr]^{-1} \times \nonumber \\ \nonumber 
& \Bigg\{ \bigg\{ 1 - \nu \s \alp \s [\q] \bigg\} \gs(u) - \nu \s  \p \s\bigg\{ [\bet] \s \sum_{k= 0}^{ u+1} (f*g)(k)\s \gs(u+1-k) \nonumber \\ \nonumber 
&+ \alp \s  \sum_{k= 0}^{ u} (f*g)(k) \s \gs(u-k) \bigg\} \nonumber \\ \nonumber 
& -  [\nu] \s [\p]^{2} \s  [\oneminustht]  \s \bigg\{ \dfrac{ \alp }{\p \s [\oneminustht]}\s \mathbb{E}[{\wx}_{X+\yet Y}(u)] +\dfrac{[\bet]}{\p \s [\oneminustht] }  \s \mathbb{E}[{\wx}_{X+\yet Y}(u+1)] \nonumber \\ \nonumber 
& - 2 \s [\nu] \s  \alp \s [\bet]  \s \sum_{k= 0}^{ u+1} (f*g)(k)\s \mathbb{E}[{\wx}_{X + WY}(u+1-k)] - \nu  [\alp]^{2} \s  \sum_{k= 0}^{ u} (f*g)(k)\s \mathbb{E}[{\wx}_{X + WY}(u-k)] \nonumber \\ \nonumber
&- [\nu] \s  [\bet]^{2}  \s \sum_{k= 0}^{ u +2} (f*g)(k)\s \mathbb{E}[{\wx}_{X + WY}(u+2-k)] \bigg\} \nonumber \\ \nonumber 
& - [\nu]^{2} \s  \p \s [\oneminustht] \s \bigg\{  [\alp]^{2} \s \sum_{k= 0}^{ u } f(k)\s \mathbb{E}[{\wx}_{X + WY}(u-k)] 
+ [\bet]^{2} \s \sum_{k= 0}^{ u+2 } f(k)\s  \mathbb{E}[{\wx}_{\si \left( X + \yet Y\right)+ \hat{Y}} (u+2-k) \nonumber \\ \nonumber 
&+ 2  \s \alp \s [\bet] \s \sum_{k= 0}^{ u+1 } f(k)\s  \mathbb{E}[{\wx}_{\si \left( X + \yet Y\right)+ \hat{Y}} (u+1-k) \bigg\} \nonumber \\ \nonumber 
& - [\nu]^{2} \s [\bet] \s [\q] \s [\oneminustht] \s \p \s \bigg\{ [\bet] \s  (f*g)(u+2) + \alp \s  (f*g)(u+1) \bigg\}  \s \gs(0)\nonumber \\ \nonumber
& - [\nu]^{2} [\p]^{2} \s  [\oneminustht] \s [\bet]  \bigg\{  [\bet] \s (f*g)(u+2)  +  \alp \s (f*g)(u+1)  \bigg\} \s  \mathbb{E}[{\wx}_{X+\yet Y}(0)] \\
& + [\nu]^{2}  \p \s  [\oneminustht] \s [\bet] \bigg\{ [\bet] \s f(u+2)  +  \alp  \s f(u+1) \s   \bigg\}  \mathbb{E}[{\wx}_{\si \left( X + \yet Y\right)+ \hat{Y}} (0) ]\Bigg\}
\end{align} 
}
for $u = 0,1,2,\ldots $ and $\gs(0)$ as given by $(\ref{initial value mzero})$. This is the first main result claimed in this paper.
 \subsection{Recursive expression for the Gerber-Shiu function when discount factor $d > 0$}
Let  $d > 0$ be the positive dividend threshold assuming non negative integer values. In this section, we derive a recursive relation for the Gerber-Shiu function corresponding to the DTSP $(\ref{model0})$  and ADTSP $(\ref{model2})$ whenever $\mathbb{S}_{0} \geq d$.
For a positive integer $d > 0$ we consider  $\gs_{d}(u)$ and $\gs_{d}^{aux}(u)$  satisfying the constraints given below $\colon$ \\
 1. Both $\mathbb{S}_{n}(u)$ and  $\mathbb{S}_{n}^{aux}(u)$ surplus processes do not pay dividends in the first time period for $u = 0,1,2,\ldots,d-1 $. \\
 2. For $u = 0,1,2,\ldots ,d$ , the dividend may or may not  be issued in the first time period.
{\footnotesize
\begin{align} \label{thm2part1}                        
&\biggl[1-v \s [ \q ] \s  \alp \s \biggr] \s \gs_{d}(u) - v \s [ \q ] \s [ \bet ] \s \gs_{d}(u+1) \nonumber \\ \nonumber
&= v \s  \p \s \tht  \s \Biggl[ [\bet] \s (\gs_{d} * f * g )(u+1) 
+  \alp \s ( \gs_{d} * f * g )(u) \Biggr] \\ \nonumber
&+ v \s \p \s [ \oneminustht ] \s\Biggl[ [ \bet ] \s ({\gs_{d}}^{aux} * f )(u+1)+  \alp \s ( {\gs_{d}}^{aux} * f )(u) \\ 
&+ v \s \p  \s [ \bet ] \s \mathbb{E} [ {\wx}_{X+\yet Y} (u+1)]  + v \s \p \s \alp \s \mathbb{E} [{\wx}_{X+\yet Y} (u)] 
\end{align} 
}
and,
{ \footnotesize
\begin{align} \label{thm2part2}
& {\gs_{d}}^{aux}(u) = v \s [ \q ] \biggl[ [\bet] \s (\gs_{d} * g)(u+1) + \alp \s (\gs_{d} * g)(u)\biggr] \nonumber \\
&+ v \s \p \s \tht  \biggl[ [\bet] \s (\gs_{d} * f * g * g)(u+1) + \alp \s (\gs_{d} * f * g * g)(u)\biggr] \nonumber \\
&+ v \s \p \s  [\oneminustht]  \biggl[ [\bet] \s ({\gs_{d}}^{aux} * f * g)(u+1) + \alp \s ({\gs}_{d}^{aux} * f * g)(u)\biggr] \nonumber \\
&+ v \s \biggl[ [\bet] \s \mathbb{E}[{\wx}_{\left( X + \yet Y\right)+ \hat{Y}} (u+1)] + \alp \s \mathbb{E}[ {\wx}_{\si{\left( X + \yet Y\right)+ \hat{Y}} }(u)]\biggr]
\end{align} 
} 
Proceeding in a manner similar to Section 2, we use the generating function technique as in $(\ref{main result})$ and compare the coefficients of $z^{u+1}$ to get a recursive relation for $\gs_{d}^{aux}(u+1)$, with $u \geq d$ as $\colon$ 
{ \footnotesize
\begin{align}\label{secnd main result}
& \gs_{d}(u+1) = \biggl[ \nu  \s [\q] \s [\bet] \biggr]^{-1} \times \nonumber \\ \nonumber 
& \Bigg\{ \bigg\{ 1 - \nu \s \alp \s [\q] \bigg\} \gs_{d}(u) - \nu \s  \p \s\bigg\{ [\bet] \s \sum_{k= 2}^{ u+1} (f*g)(k)\s \gs_{d}(u+1-k) \nonumber \\ \nonumber 
&+ \alp \s  \sum_{k= 2}^{ u} (f*g)(k) \s \gs_{d}(u-k) \bigg\} \nonumber \\ \nonumber 
& -  [\nu] \s [\p]^{2} \s  [\oneminustht]  \s \bigg\{ \dfrac{ \alp }{\p \s [\oneminustht]}\s \mathbb{E}[{\wx}_{X+\yet Y}(u)] +\dfrac{[\bet]}{\p \s [\oneminustht] }  \s \mathbb{E}[{\wx}_{X+\yet Y}(u+1)] \nonumber \\ \nonumber 
& - 2 \s [\nu] \s  \alp \s [\bet]  \s \sum_{k= 2}^{ u+1} (f*g)(k)\s \mathbb{E}[{\wx}_{X+\yet Y}(u+1-k)] - \nu  [\alp]^{2} \s  \sum_{k= 2}^{ u} (f*g)(k)\s \mathbb{E}[{\wx}_{X+\yet Y}(u-k)] \nonumber \\ \nonumber
&- [\nu] \s  [\bet]^{2}  \s \sum_{k= 2}^{ u +2} (f*g)(k)\s \mathbb{E}[{\wx}_{X+\yet Y}(u+2-k)] \bigg\} \nonumber \\ \nonumber 
& - [\nu]^{2} \s  \p \s [\oneminustht] \s \bigg\{  [\alp]^{2} \s \sum_{k= 1}^{ u } f(k)\s \mathbb{E}[{\wx}_{X+\yet Y}(u-k)] 
+ [\bet]^{2} \s \sum_{k= 1}^{ u+2 } f(k)\s  \mathbb{E}[{\wx}_{\si \left( X + \yet Y\right)+ \hat{Y}} (u+2-k) \nonumber \\ \nonumber 
&+ 2  \s \alp \s [\bet] \s \sum_{k= 1}^{ u+1 } f(k)\s  \mathbb{E}[{\wx}_{\si \left( X + \yet Y\right)+ \hat{Y}} (u+1-k) \bigg\} \nonumber \\ \nonumber 
& - [\nu]^{2} \s [\bet] \s [\q] \s [\oneminustht] \s \p \s \bigg\{ [\bet] \s  (f*g)(u+2) + \alp \s  (f*g)(u+1) \bigg\}  \s \gs(0)\nonumber \\ \nonumber
& - [\nu]^{2} [\p]^{2} \s  [\oneminustht] \s [\bet]  \bigg\{  [\bet] \s (f*g)(u+2)  +  \alp \s (f*g)(u+1)  \bigg\} \s  \mathbb{E}[{\wx}_{X+\yet Y}(0)] \\
& + [\nu]^{2}  \p \s  [\oneminustht] \s [\bet] \bigg\{ [\bet] \s f(u+2)  +  \alp  \s f(u+1) \s   \bigg\}  \mathbb{E}[{\wx}_{\si \left( X + \yet Y\right)+ \hat{Y}} (0) ]\Bigg\}. 
\end{align} 
}

Since $\mathbb{S}_{0} \geq d$, the initial values for the above expression when
$u = 0,1,2,\ldots ,d$ are to be determined.
Let $d=0$  so that the initial values may be found. Consider the  penalty function, 
$ \wa(v_{1},v_{2})= \mathbb{I}{\lbrace x=v_{1},y=v_{2}\rbrace}$ with $x = 0, 1, 2, 3, \ldots$ and $y = 1, 2, 3, \ldots$ and where
$v_{1}\in \lbrace 0,1, 2, \cdots \rbrace$ and  $v_{2}\in \lbrace1, 2, 3, \cdots \rbrace$ are constants.
The discounted joint probability mass function corresponding to $\mathbb{S}_{\tau -}$ and $ |\mathbb{S}_{\tau}(u)|$ and $\mathbb{S}_{0}=d=0$ is given by :
\begin{align} \label{muv1v2}
\mu(v_{1},v_{2})= \mathbb{E}\left[v^{\tau} \mathbb{I} \s \lbrace \mathbb{S}_{\tau -}= v_{1},| \mathbb{S}_{\tau}| = v_{2} \rbrace \mathbb{I} {\lbrace \tau < \infty \rbrace} \s |\s \mathbb{S}_{0}=0 \right]
\end{align} 
 $v_{1}\in \lbrace 0,1, 2, \cdots \rbrace \text{and} v_{2}\in \lbrace1, 2, 3, \cdots \rbrace$
With the introduction of the penalty function $(\ref{muv1v2})$, it can be easily shown that  $\mathbb{E}[{\wx}_{X}(u)] = \mathbb{I}{\lbrace u=v_{1}\rbrace} P(X=u+v_{2}) $,  $\mathbb{E}[{\wx}_{X}(0)] = \mathbb{I}{\lbrace v_{1}=0 \rbrace} f(v_{2}) $,  $\mathbb{E}[{\wx}_{X + \yet Y}(0)] =  \mathbb{I}{\lbrace v_{1}=0 \rbrace} { [\tht] (f*g)(v_{2})+[ \oneminustht] f(v_{2})} $ \\ 
and $\mathbb{E}[{\wx}_{\si(X + \yet Y)+\hat{Y}}(0)] = \mathbb{I}{\lbrace v_{1}=0 \rbrace} \biggl[  [\q] g(v_{2}) + \p [ \oneminustht] \s (f * g)(v_{2}) + \p \s [\tht] (f * g * g)(v_{2})\biggr]$ 
Also, when $v_{1}\neq 0$ , we have,
$\mathbb{E}[{\widetilde{\wx}}_{X}(z)] = \sum_{u=0}^{\infty}z^{u}{\wx}_{X}(u) = \sum_{u=0}^{\infty}z^{u} \mathbb{I}{\lbrace v_{1}=u\rbrace}P(X=u+v_{2}) $ \\
Substituting the above results in addition to $(\ref{muv1v2})$ in $(\ref{initial value mzero})$ and simplifying, we can find the initial values $\gs_{d}(0),\ldots,\gs_{d}(d)$. $\mu(0,v_{2}) $ may be obtained by replacing the penalty function in 
 in expression $(\ref{initial value mzero})$ by the  penalty function $\wa(v_{1},v_{2})= \mathbb{I} \lbrace{x=v_{1},y=v_{2} \rbrace}$. 
 
{ \footnotesize
\begin{align} \label{mu zero v2}
&\mu(0,v_{2}) = \biggl[ \dfrac{\p}{\q} \biggr]  \bigg[  \dfrac{\alp \s z_{0}}{[\bet]}  \biggr] \bigg\{ \tht \s ( f * g ) (v_{2}) + [\oneminustht] f(v_{2}) \bigg\} \nonumber \\ \nonumber
& + \Bigg\{  \biggl[ \dfrac{\p}{\q} \biggr] \biggl[  \dfrac{\nu \s [\oneminustht] \s [\bet]}{\az(z_{0}} \biggr] \ftilde(z_{0}) \s  \bigg[ 1+  \dfrac{\alp \s z_{0}}{[\bet]}  \biggr] \Bigg\} \s \times \nonumber \\ 
& \bigg[  \dfrac{\alp \s z_{0}}{[\bet]}  \biggr] \s \biggl[ [\q] \s g(v_{2}) + \p \s [\oneminustht] \s (f*g)(v_{2})   +  \p \s \tht \s (f*g*g)(v_{2})  \biggr] 
\end{align} 
}
On similar lines, for $v_{1}, v_{2}\in \mathbb{N}$, we obtain,
{ \footnotesize
\begin{align} \label{explicit mu v1 v2}
&\mu(v_{1},v_{2}) =  \biggl[ \dfrac{\p \s {z_{0}}^{v_{1}}}{\q} \biggr] \biggl[ 1 +  \dfrac{\alp \s z_{0}}{[\bet]}  \biggr] \biggl[ \tht \s ( f * g ) (v_{1}+v_{2}) + [\oneminustht] f(v_{1}+v_{2}) \biggr] \nonumber \\ \nonumber
 & + \biggl[ \dfrac{\p}{\q} \biggr] \biggl[  \dfrac{\nu \s [\oneminustht] \s [\bet]}{\az(z_{0})} \biggr]  \biggl[ 1 +  \dfrac{\alp \s z_{0}}{[\bet]} \biggr]^{2}  {z_{0}}^{v_{1}} \ftilde(z_{0}) \times \nonumber \\ 
&\biggl[ [\q] \s g(v_{1}+v_{2}) + (f*g)(v_{1}+v_{2}) \s \p \s [\oneminustht] + (f*g*g)(v_{1}+v_{2}) \s \p \s \tht \biggr]
\end{align}
}
Since dividend is not issued in the first time period, $P(V=1)=0$.  Hence, setting $\mathbb{E}(V) = 0$ in $(\ref{thm2part1})$ and $(\ref{thm2part2})$ will respectively yield $\colon$
{ \footnotesize
\begin{align} \label{E(V)=0 md(u)}
&\gs_{d}(u) = \nu \s [\q] \s \gs_{d}(u+1) + \nu \s \p \s \mathbb{E}[{\wx}_{X+WY}(u+1)] \nonumber \\ 
&+ \nu \s \p \biggl[ \tht \s  \sum_{k= 1}^{ u+1}  \s \gs_{d}(u+1-k) (f*g)(k) + [\oneminustht] \s \sum_{k= 1}^{ u+1}  \s {\gs}^{aux}_{d}(u+1-k) (f)(k) \biggr]
\end{align}
}
and, 
{ \footnotesize
\begin{align}  \label{E(V)=0 md aux(u)}
&{\gs}^{aux}_{d}(u) = \nu \Bigg\{  \mathbb{E}[{\wx}_{K(X+WY)+\hat{Y}}(u+1)] +  [\q] \s \sum_{k= 1}^{ u+1}  \s \gs_{d}(u+1-k) \s (g)(k) \nonumber \\ 
&+  \p \s [\oneminustht] \biggl[ \sum_{k= 2}^{ u+1}  \s {\gs}^{aux}_{d}(u+1-k) \s (f*g)(k) + \sum_{k= 3}^{ u+1}  \s \gs_{d}(u+1-k) (f*g*g)(k) \biggr] \Bigg\}
\end{align}
}
The first set of $2d$ equations to determine the $2d+1$ unknowns $\gs_{d}(0),\ldots,\gs_{d}(d)$ and $\gs_{d}^{aux}(0),\ldots,\gs_{d}^{aux}(d-1)$ are obtained by setting $u = 0,1,2,\ldots,d-1$ in $(\ref{E(V)=0 md(u)})$ and $(\ref{E(V)=0 md aux(u)})$. 
Using expression $(\ref{explicit mu v1 v2})$, the joint probability mass function corresponding to the surplus immediately before ruin and the deficit at ruin we can compute $\gs_{d}(d)$, as the marginal distribution which is given by $\colon$
{ \footnotesize
\begin{align} \label{mdd}
{\gs}_{d}(d) = \mathop{\sum_{v_{1}=0}^{\infty}\sum_{v_{2}=1}^{d}} \mu(v_{1},v_{2}) \gs_{d}(d - v_{2}) + \mathop{\sum_{v_{1}=0}^{\infty}\sum_{v_{2} = d+1}^{\infty}} \mu(v_{1},v_{2}) \wa ( d + v_{1} , d-v_{2} ).
\end{align}
}
\section{Applications}
Recursive expressions for a few ruin related quantities of interest will be derived in this section.
Let $F(n) = \sum_{k=1}^{n}{P}(X=k)$ and $\overline{F}(n) = 1 - \sum_{k=1}^{n}{P}(X=k)$. \\
In Example $(\ref{firstexample})$, a recursive expression for the probability of ruin $\phi(u)$ is obtained. 
\begin{example} \label{firstexample}
Let $\wa(x, y) = 1$. Then $\gs(u) =\nu \s  P[\tau < \infty|\mathbb{S}_{0} = u]=\phi(u)$ = probability of ruin. 
\end{example}
Here, $ \mathbb{E}[{\wx}_{X+WY}(u)]= \overline{F}(u)$, $ \mathbb{E}[{\wx}_{X+WY}(0)]= 1$,    $\mathbb{E}[{\wx}_{K(X+WY)+\hat{Y}}(u)] = 1 - P(u) = \overline{F}(u)$ \\
and $\mathbb{E}[{\wx}_{K(X+WY)+\hat{Y}}(0)] = 1$. Further, $\mathbb{E}[\tilde{\wx}_{X+WY}(z_{0})] = \sum_{u= 0}^{ \infty}  {z_{0}}^{u} [\overline{F}(u)] $.  A recursive expression for the probability of ruin is obtained by substituting the quantities mentioned in this example and replacing $\gs(u)$ by $\phi(u)$ in $(\ref{main result})$.  The initial value $\gs(0)$ may be obtained from 
$(\ref{initial value mzero})$. \\
In Example $(\ref{secondexample})$, a recursive formula for the probability of deficit at ruin is obtained.
\begin{example} \label{secondexample}
Let $\wa(x_{1}, x_{2})= \mathbb{I}\lbrace x_{2} = y \rbrace$, $y = 1, 2, 3, \ldots$. Then, \\
$\gs(u) = {P}\biggl[|\mathbb{S}_{\tau_{-}}| = y | \mathbb{S}_{0}=u\biggr]$ = Probability of the deficit at ruin = $G(u,y)$ (say)
\end{example}
Here, for any $y = 1, 2, 3, \cdots $, $ \mathbb{E}[{\wx}_{X+WY}(u)]= P(X + WY = u +y)$, $ \mathbb{E}[{\wx}_{X+WY}(0)]= P(X + WY = y)$,    $\mathbb{E}[{\wx}_{K(X+WY)+\hat{Y}}(u)] = P[{K(X+WY)+\hat{Y}}= u+y]$ and $\mathbb{E}[{\wx}_{K(X+WY)+\hat{Y}}(0)] =  P[{K(X+WY)+\hat{Y}}= y]$. Further, $\mathbb{E}[\tilde{\wx}_{X+WY}(z_{0})] = \sum_{u= 0}^{ \infty}  {z_{0}}^{u}P(X + WY = u +y)$ and $\mathbb{E}[{\wx}_{K(X+WY)+\hat{Y}}(z_{0}]= \sum_{u= 0}^{ \infty}  {z_{0}}^{u} P[{K(X+WY)+\hat{Y}}= u+y]  $.  A recursive expression for the probability of deficit at ruin is obtained  by substituting the above quantities and replacing $\gs(u)$ by $G(u,y)$ in $(\ref{main result})$.  The initial value $\gs(0)$ may be obtained from $(\ref{initial value mzero})$. \\
In Example $(\ref{thirdexample})$, a recursive expression for the generating function of the deficit at ruin is obtained. 
\begin{example} \label{thirdexample}
Let $\wa(x_{1}, x_{2}) = r^{x_{2}}$ and  then, 
$\gs(u) = \nu \mathbb{E} [r^{|S_{\tau -} \s |} \mathbb{I}\lbrace \tau < \infty \rbrace] = \widehat{G}(u,r )$ (say) , the generating function of the deficit at ruin.
\end{example}
Here, $\mathbb{E}[{\wx}_{X+WY}(u)] = \sum_{k=1}^{\infty} r^{k}P(X+WY=k+u)$, $\mathbb{E}[{\wx}_{K(X+WY)+\hat{Y}}(u)] =  \sum_{k=1}^{\infty} r^{k}P[{K(X+WY)+\hat{Y}}= k+u]$, $\mathbb{E}[{\wx}_{X+WY}(0)] = \sum_{k=1}^{\infty} r^{k}P(X+WY=k)$ and $\mathbb{E}[{\wx}_{K(X+WY)+\hat{Y}}(0)] =  \sum_{k=1}^{\infty} r^{k}P[{K(X+WY)+\hat{Y}}= k]$.
 Further, $\mathbb{E}[\tilde{\wx}_{X+WY}(z_{0})] = \mathop{\sum_{u=0}^{\infty}\sum_{k=1}^{\infty}}{z_{0}}^{u}P(X + WY = z_{0} +k)$ and $\mathbb{E}[{\wx}_{K(X+WY)+\hat{Y}}(z_{0})] = \mathop{\sum_{u=0}^{\infty}\sum_{k=1}^{\infty}}{z_{0}}^{u}P(K(X+WY)+\hat{Y} = z_{0} + k)$.  \\ 
 A recursive expression for the generating function of the deficit at ruin is obtained by substituting the above quantities and replacing $\gs(u)$ by $\widehat{G}(u,r )$ in $(\ref{main result})$ .The initial value $\gs(0)$ may be obtained from $(\ref{initial value mzero})$.  \\
 In Example $(\ref{fourthexample})$, a recursive expression for the probability of the surplus at ruin is obtained. 
 \begin{example} \label{fourthexample}
Let ${\wa}(x_{1},x_{2}) = \mathbb{I}\lbrace x_{1} = y \rbrace$, $y = 1, 2, 3, \ldots$ then $\gs(u) = \nu \s  \mathbb{E} [\mathbb{I}\lbrace \mathbb{S}_{\tau -} =y \rbrace\mathbb{I}\lbrace \tau < \infty \rbrace| \mathbb{S}_{0} = u]= {P}(\mathbb{S}_{\tau -} = y)$ = Probability of the surplus at ruin = $s(u,y)$  (say)
\end{example}
 Here $\mathbb{E}[{\wx}_{X+WY}(u)] = \nu \s \mathbb{I}\lbrace u = y \rbrace \s  \overline{F}(u)$, $\mathbb{E}[{\wx}_{K(X+WY)+\hat{Y}}(u)] = \mathbb{I} \lbrace u = y \rbrace \s  \overline{F}(u),  $ and   $\mathbb{E}[{\wx}_{X+WY}(0)] = \mathbb{I} \lbrace  y = 0 \rbrace \s  \bar {F}(u) = 0$ and $\mathbb{E}[{\wx}_{K(X+WY)+\hat{Y}}(0)] = \mathbb{I} \lbrace y = 0 \rbrace \s \overline{F}(u) = 0$. Further, $\mathbb{E}[\tilde{\wx}_{X+WY}(z_{0})] = \sum_{u=0}^{\infty}{z_{0}}^{u} \mathbb{I}\lbrace z_{0} = y \rbrace \overline{F}(z_{0})$  and  $\mathbb{E}[{\wx}_{K(X+WY)+\hat{Y}}(z_{0})] = \sum_{u=0}^{\infty}{z_{0}}^{u} \mathbb{I}\lbrace z_{0} = y \rbrace \overline{F}(z_{0})$. \\
 A recursive expression for the probability of the surplus at ruin is obtained  by substituting the above quantities and replacing $\gs(u)$ by $s(u,y)$ in $(\ref{main result})$.  The initial value $\gs(0)$ may be obtained from $(\ref{initial value mzero})$.  
 \begin{example} \label{fifthexample}
 Let ${\wa}(x_{1},x_{2}) = \mathbb{I}\lbrace x_{1}+x_{2} = y \rbrace$, $y = 1, 2, 3, \ldots$ then
  $\gs(u) = \nu \s \mathbb{E} [\mathbb{I}\lbrace \mathbb{S}_{\tau -} +| \mathbb{S}_{\tau}|  = y\rbrace| \mathbb{S}_{0} = u]$ = $\nu$ (Probability of the claim causing ruin) = $\nu \s l(u,y)$
 \end{example}
 Here, $\mathbb{E}[{\wx}_{X+WY}(u)] = \sum_{k=1}^{\infty} \mathbb{I} \lbrace k = y \rbrace P(X+WY=k)$ = $P(X+WY = y)$, $\mathbb{E}[{\wx}_{K(X+WY)+\hat{Y}}(u)] =P(K(X+WY)+\hat{Y} = y)$, $\mathbb{E}[{\wx}_{X+WY}(0)] = 0 $, $\mathbb{E}[{\wx}_{K(X+WY)+\hat{Y}}(0)] = 0$, $\mathbb{E}[\tilde{\wx}_{X+WY}(z_{0})] = \sum_{u=0}^{\infty}{z_{0}}^{u} P(X+WY = y)$ and $\mathbb{E}[{\wx}_{K(X+WY)+\hat{Y}}(z_{0})] = \sum_{u=0}^{\infty}{z_{0}}^{u} P(K(X+WY)+\hat{Y} = y)$. \\
  A recursive expression for the probability of the claim causing ruin is obtained  by substituting the above quantities and replacing $\gs(u)$ by $\nu \s l(u,y)$ in $(\ref{main result})$.  The initial value $\gs(0)$ may be obtained from $(\ref{initial value mzero})$. 
\section{Conclusion}
In this paper, a recursive expression for the conditional expected penalty function for the  risk model under consideration has been obtained under the assumption that the probabilities of occurrence of claim , by-claim and probability the issuance of dividends follow a Beta distribution. However, in literature, to the best of our knowledge, the probabilities are assumed to be fixed (constant). The results obtained in this paper are generalizations of the standard results obtained for the Compound Binomial risk model with by-claims and randomized dividends. If the probabilities are fixed, then the results in \cite{wat2018compound} and \cite{yuen2013discrete} can be reproduced. Recursive expressions for ruin related quantities i.e. the probability of ruin, probability of the deficit at ruin, generating function for the deficit at ruin and probability of the surplus at ruin has been arrived at.
 
\bibliography{references}
\bibliographystyle{plain}
\end{document}